\documentclass[a4paper,10pt,journal]{IEEEtran}

\ifCLASSINFOpdf
\else
\fi

\usepackage{algorithm,algorithmic}
\usepackage[T1]{fontenc}
\usepackage[utf8]{inputenc}
\usepackage{newunicodechar}
\usepackage{booktabs}
\usepackage{lipsum}
\usepackage{fullpage}
\usepackage{graphicx}
\usepackage{epstopdf}
\usepackage{float}
\usepackage{tablefootnote}
\usepackage{threeparttable}
\usepackage[font=footnotesize]{caption}
\usepackage{pstricks}
\usepackage{color}
\usepackage{subfig}
\usepackage{cite}
\usepackage{amsmath}
\usepackage{multirow}

\newunicodechar{→}{$\to$}
\graphicspath{{./figs/}}

\begin{document}

%
\title{DASH Adaptation Algorithm based on Adaptive Forgetting Factor estimation}
%
%
%

\author{Miguel~Aguayo,
        Luis~Bellido,
        Carlos~M.~Lentisco,
        and~Encarna~Pastor
\thanks{This work was supported in part by the Spanish Ministry of Economy and Competitiveness in the context of the project GREDOS, reference TEC2015-67834-R (MINECO/FEDER, UE), and the project Elastic Networks, reference TEC2015-71932-REDT.}
\thanks{The authors are with the Department
of Telematics Systems Engineering, Universidad Politécnica de Madrid, Madrid 28040, Spain (e-mail: aguayo@dit.upm.es; lbellido@dit.upm.es; clentisco@dit.upm.es; epastor@dit.upm.es).}
\thanks{This is the authors’ accepted version of the article. The final version published by IEEE is M. Aguayo, L. Bellido, C. M. Lentisco and E. Pastor, "DASH Adaptation Algorithm Based on Adaptive Forgetting Factor Estimation," in IEEE Transactions on Multimedia, vol. 20, no. 5, pp. 1224-1232, May 2018, doi: 10.1109/TMM.2017.2764325.}
\thanks{©2018 IEEE. Personal use of this material is permitted. Permission from IEEE must be obtained for all other uses, in any current or future media, including reprinting/republishing this material for advertising or promotional purposes, creating new or redistribution to servers or lists, or reuse of any copyrighted component of this work in other work.}}
\maketitle

\begin{abstract}
The wide adoption of multimedia service-capable mobile devices, the availability of better networks with higher bandwidths, and the availability of platforms offering digital content has led to an increasing popularity of multimedia streaming services. However, multimedia streaming services can be subject to different factors that affect the quality perceived by the users, such as service interruptions or quality oscillations due to changing network conditions, particularly in mobile networks. Dynamic Adaptive Streaming over HTTP (DASH), leverages the use of content-distribution networks and the capabilities of the multimedia devices to allow multimedia players to dynamically adapt the quality of the media streaming to the available bandwidth and the device characteristics. While many elements of DASH are standardized, the algorithms providing the dynamic adaptation of the streaming are not. The adaptation is often based on the estimation of the throughput or a buffer control mechanism. In this paper, we present a new throughput estimation adaptation algorithm based on a statistical method named Adaptive Forgetting Factor (AFF). Using this method, the adaptation logic is able to react appropriately to the different conditions of different types of networks. A set of experiments with different traffic profiles show that the proposed algorithm improves video quality performance in both wired and wireless environments.
\end{abstract}


%
\IEEEpeerreviewmaketitle

\vspace{0.5cm}
\section{Introduction}
%
%
%
%
\IEEEPARstart{T}{he} Internet traffic nowadays is mostly real-time entertainment traffic (audio and video). A recent Internet usage report~\cite{sandvine}~shows that real-time entertainment traffic consumes 65.35 percent of the Internet backbone aggregate traffic. Video streaming traffic is expected to experiment a growth of 67 percent in mobile and 29 percent in fixed networks \cite{cisco} in the future. The explosion of multimedia content has driven the industry and research community to create protocols and architectures to deliver video services to all users. The 3\textsuperscript{rd} Generation Partnership Project (3GPP) has defined the use of Dynamic Adaptive Streaming over HTTP (DASH)~\cite{3gpp}~as the standard for multimedia delivery in mobile networks, specifically in Long Term Evolution (LTE) networks. In a previous work \cite{lentisco2} we have proposed solutions to improve the delivery of a DASH-encoded multimedia content that is broadcast over LTE. In this paper, the focus is on multimedia unicast services based on DASH.\par

One of the advantages of using HTTP is that standard HTTP servers and content distribution techniques can be reused for storing and delivering multimedia content. The DASH standard defines how a multimedia content can be divided into small files or “chunks”, and how to store the description of chunks in a metadata file named Media Presentation Description (MPD) so a multimedia player can retrieve both metadata and the sequence of chunks using HTTP to play the multimedia content.\par

Different representations make it possible that the same content can be retrieved with different qualities, depending on the user terminal or on the available bandwidth. For each representation the MPD file contains information about media type, codec, video width and height, frame rate, average bitrate, and so on. The actual media files, or media segments, are identified by Uniform Resource Locators (URL). \par

A DASH player implements an adaptation algorithm monitoring network conditions (bandwidth, delay, and so on) and selecting the representation to be downloaded. The aim of the adaptation algorithm is to ensure that the client selects a representation with the most appropriate bitrate to obtain the highest quality, while avoiding stalling events during the media playback. A guideline with remarks on possible client behavior is provided as an annex in~\cite{iso} but adaptation algorithms are not standardized.\par

Substantial research exists in adaptive algorithms for DASH. Many of the algorithms have been designed for specific scenarios, e.g., mobile wireless networks, so to establish the parameters to achieve the highest quality allowed by the available network conditions. However, the same adaptation algorithm in a different scenario can overestimate or underestimate the available network bandwidth. When an adaptation algorithm is estimating more bandwidth than the network is offering, video stalling happens, because of buffer under-runs. On the other hand, if an adaptation algorithm underestimates the network bandwidth, the video player retrieves video qualities which are lower than what the network conditions permit, hence affecting the quality perceived by the user.\par

In this paper, we propose to use the Adaptive Forgetting Factor (AFF) method to improve throughput estimation in DASH adaptation algorithms. Our proposal is based on the capability of AFF to quickly adapt to short-term fluctuations of the bandwidth, especially in wireless networks. Using AFF as a throughput estimation technique, a DASH multimedia player will be able to achieve better quality in the video playback in both wired and wireless scenarios.\par

The rest of the paper is organized as follows. Section II describes the state of the art on adaptation algorithms. Section III presents the adaptation algorithm based on AFF. Section IV describes the scenario used to test the proposed adaptation algorithm. Section V presents the obtained results. Section VI presents a fairness analysis of the AFF algorithm. Finally, section VII presents the conclusions and future work. \par

\section{Overview on Adaptation Algorithms}
A DASH client uses an adaptation algorithm to handle the selection of the multimedia representation for each segment that needs to be downloaded. This selection is based on the network conditions, which are compared to a set of parameters on the client (e.g., buffer level) to make the choices about the highest possible quality when requesting the next media segment. This process of adaptation can be assisted by intermediate network nodes which have information on how much bandwidth is going to be allocated to the clients \cite{lentisco4,essaili,bouten}. But, in general, adaptation algorithms implement buffer control and throughput estimation methods. Buffer control is designed so that the fluctuations of the bandwidth, especially in wireless environments, will not affect the playback of the video. Throughput estimation is designed to maximize the quality in terms of bitrate selection by estimating the available bandwidth. This paper focuses on throughput estimation algorithms.\par

Throughput estimation algorithms focus on providing an adequate estimation of the throughput that the client can obtain from the network. 
Most of the adaptation algorithms start by calculating the instant throughput \cite{tian}, which is defined as the size of the last downloaded segment divided by the time taken to download it.\par

The problem with instant throughput is that it is not an appropriate throughput estimation method because measurements can fluctuate from one segment to another. Using it as a throughput estimator would have the effect of the multimedia player continuously adapting the bitrate to the instant throughput measurements, which would affect the quality of the playback. Adaptation algorithms that use instant throughput measurements as a throughput estimation method, combine it with other mechanisms. For example, BOLA \cite{bola}, adapts the video bitrate using a buffer control method.\par

Some methods for estimating the throughput rely on the measurements of lower layers \cite{oyman, oymandos}, (e.g., from the physical layer). Those measurements are then compared to the instant throughput to make an adaptation decision. However, throughput measurements from the physical layer have the inconvenience of including all network services from the client.\par

Other throughput oriented algorithms calculate the average throughput for the last N segments of video obtained by the client \cite{tian}. However, this method has the disadvantage of not detecting short-term fluctuations of the available bandwidth that can occur, e.g., in wireless networks. If a short-term decrease of available bandwidth is not detected, this would lead to buffer starvation. \par

Lin \textit{et al.} \cite{lin} proposed to use the mean, the standard deviation and the fluctuation of the throughput to estimate the bandwidth. This method has the problem of heuristically establishing the fluctuation parameter with values of 0 to 0.025 for wired networks and 0 to 0.1 for wireless networks.\par

An alternative throughput estimation method is to calculate the harmonic mean of a certain number of past measurements. The FESTIVE algorithm \cite{festive} uses the last twenty measurements while the ELASTIC algorithm \cite{elastic} uses the last five. The harmonic mean method functions optimally when having steady bandwidth measurements because it can discriminate some of the outlier measurements. However, in wireless environments, there are short-term bandwidth fluctuations that can cause this method to overestimate or underestimate the available bandwidth. \par

Zhou \textit{et al.} \cite{zhou} proposed the use of a rate adaptation algorithm based on Markov decision processes that uses the mean and a temporal variance as a mechanism for bitrate switching. The problem of continuous bitrate switching is addressed by establishing two buffer thresholds. Using an algorithm that is also based on Markov decision processes, \cite{bokani} proposes the use of a reinforcement learning method with a reward function to program new segment petitions. \par

The EWMA method is proposed in \cite{akh,thang}. EWMA behaves as a low-pass filter for the throughput measurements. EWMA applies weighting factors so the weighting of each older instant throughput measurement decreases exponentially. The problem with this method is that the initial weight parameter needs to be fixed to a different value depending on the type of network. Usually weight values of 0 to 0.1 are proposed for wired networks and 0.2 to 0.3 for wireless networks. This can cause the method to not behave adequately (e.g., producing stalling events) when the weight parameters are not set adequately for different network scenarios. \par

Li \textit{et al.} \cite{panda} proposed PANDA, an algorithm that uses EWMA with an exponential weight of 0.2 as a throughput estimator, which would be adequate for wireless networks. The algorithm also proposes the use of an Additive Increase Multiplicative Decrease (AIMD) bitrate selection algorithm and a random scheduler algorithm to avoid playback synchronization for simultaneous players. \par

Another example of an EWMA based throughput estimation method adapted to a specific type of network can be found in \cite{miller}, which proposed the use of DASH in a dense wireless network scenario, using proportional-integral-derivative (PID) controllers and an EWMA throughput estimation in every wireless client to manage the quality selection and client scheduling. \par

Thang \textit{et al.} \cite{le} presented a method that combines the use of EWMA and the Round Trip Time (RTT) estimation method of TCP but presents the problem of setting a fixed weight parameter depending on the type of network used. Jeong and Chung \cite{jeong} proposed to use EWMA, RTT and a Media Segment Duration (MSD) measurement, but a fixed weight parameter depending on the access network is needed, like \cite{akh,thang,le,jeong,thang2}. \par

Lai \textit{et al.} \cite{lai} proposed the use of EWMA and two correcting parameters that are used to calculate the weight used in EWMA, and a safety margin of three times the standard deviation in the EWMA formula. \par

A combination of EWMA and a dynamic fluctuation factor in \cite{kim} tries to compensate for the error between the last throughput measurement and the next. However, this method is based on comparing the last throughput measurement to the previous one, making the exponential parameter to adapt too slowly when there are bandwidth fluctuations. \par

The AFF method proposed in this paper to address the problem of bandwidth fluctuations is explained in the following section. \par

\section{Adaptive Forgetting Factor for DASH}

The AFF method was originally designed as a recursive least-square adaptive filter to recover data from corrupted signals \cite{aff}. Similar methods can be found in various fields like medicine, finance, computer networks monitoring or astronomy.\par

In this paper we propose the use of an AFF method originally designed by Bodenham \cite{bodenhama} as a throughput estimation method for network security. This method is based on statistical process control to analyze streams of data and detect changes using the mean and the variance.\par

AFF shares with EWMA the idea of using a weight parameter that decreases the impact of older measurements exponentially (functioning as a smoothing parameter) on the estimation of the mean. But while EWMA uses a fixed weight factor, AFF calculates the value dynamically. This allows AFF to react quicker to short-term fluctuations of the bandwidth. The estimated throughput in the AFF method is shown in (1). The AFF mechanism is $\overrightarrow{\lambda}$, where
$\overrightarrow{\lambda}$ = ($\lambda_{0}$, $\lambda_{1}$,...,$\lambda_{N}$) and $\lambda_{i}$  $\in$ (0,1).\par

\begin{equation}
\label{eq1}
 \overline{THR}_{N,\overrightarrow{\lambda}}= \frac{m_{N,\overrightarrow{\lambda}}}{w_{{N,\overrightarrow{\lambda}}}}~~~~~~N \geq 1
\end{equation}

\begin{equation}
\label{eq2}
 m_{N,\overrightarrow{\lambda}}=\lambda_{N-1}m_{N-1,\overrightarrow{\lambda}}+THR~~~~N \geq 1
\end{equation}
 
 \begin{equation}
\label{eq3}
 w_{N,\overrightarrow{\lambda}}=\lambda_{N-1}w_{N-1,\overrightarrow{\lambda}}+1~~~~N \geq 1
\end{equation}\par

Equation (\ref{eq1}) is the throughput estimation, (\ref{eq2}) represents the accumulated instant throughput measurements. (\ref{eq3}) shows the accumulated of the number of segments. Where $m_{0,\overrightarrow{\lambda}}=0$ , $w_{0,\overrightarrow{\lambda}}=0$ and $\lambda_{0}=1$ respectively. \par

As shown in (\ref{eq2}) and (\ref{eq3}), the key component of the AFF method is how the weight factor is updated, namely $\lambda_N\rightarrow\lambda_{N+1}$. To obtain $\lambda_N$, it is necessary to apply an online optimization to minimize a cost function $L_{N+1,\overrightarrow{\lambda}}$.This is possible by applying a first order optimization algorithm such as the one-step gradient descent (\ref{eq4}). \par
\begin{equation}
 \label{eq4}
 \lambda_{N+1}=\lambda_{N}-\eta\frac{\partial}{\partial \overrightarrow{\lambda}}L_{N+1,\overrightarrow{\lambda}}
\end{equation}

 \begin{equation}
\label{eq5}
L_{N+1,\overrightarrow{\lambda}}=\left[ \overline{THR}_{N,\overrightarrow{\lambda}}-THR_{N+1}\right]^2
\end{equation} \par
In (\ref{eq4}) $\eta$ is the step size and $\eta\ll1$ so that the algorithm can react faster when solving the optimization. The cost function, shown in (\ref{eq5}), compares the average throughput values to the new measurement to assure that the estimated value is as close as possible to the new measurement.\par

Utilizing the chain rule method in (\ref{eq5}), the result is a derivative of (\ref{eq1}), that translates in derivatives of (\ref{eq2}) and (\ref{eq3}). To solve those derivatives a differentiation from first principle (delta method) is applied. This method aims to find the instant rate of change of (\ref{eq2}) and (\ref{eq3}) with respect of $\overrightarrow{\lambda}$. \par
Equations (\ref{eq7}) and (\ref{eq8}) present the results of applying the first principle or delta method to (\ref{eq2}) and (\ref{eq3}). For lemma proof or explanations regarding the mathematical methods applied in this research consult \cite{bodenhamt}.\par

 \begin{equation}
\label{eq6}
 \Delta_{N,\overrightarrow{\lambda}}=\lambda_{N-1}\Delta_{N-1,\overrightarrow{\lambda}}+m_{N-1,\overrightarrow{\lambda}}
\end{equation}
 \begin{equation}
\label{eq7}
 \Omega_{N,\overrightarrow{\lambda}}=\lambda_{N-1}\Omega_{N-1,\overrightarrow{\lambda}}+w_{N-1,\overrightarrow{\lambda}}
\end{equation} \par
In (\ref{eq6}) and (\ref{eq7}) the initial measurements are defined as: $\Delta_{1,\overrightarrow{\lambda}}=0$ and $\Omega_{1,\overrightarrow{\lambda}}=0$. Next we proceed to solve the derivative of (\ref{eq1}) which is shown in (\ref{eq8}). \par

 \begin{equation}
\label{eq8}
\frac{\partial}{\partial \overrightarrow{\lambda}}\overline{THR}_{N,\overrightarrow{\lambda}}= \frac{\Delta_{N,\overrightarrow{\lambda}}m_{N,\overrightarrow{\lambda}}-\Omega_{N,\overrightarrow{\lambda}}w_{N,\overrightarrow{\lambda}}}{\Delta_{N,\overrightarrow{\lambda}}}
\end{equation}\par

The equations formerly explained were implemented as shown in Algorithm 1. The aim of the AFF method is to correctly select the representation whose bitrate matches the bandwidth conditions that the network offers to achieve the highest quality in the video playback. This is achieved by estimating adaptively the throughput of the video streaming, and when encountering short-term fluctuations, placing greater weights on more recent observations and thereby “forgetting” older measurements faster. \par

The algorithm compares if the actual measurement is the first, because in that case the calculation is done with different initialization variables.
Afterwards the algorithm calculates the estimated throughput using the next instant throughput measurement in the AFF equations. \par

 \begin{algorithm}
 \label{affalg}
 \caption{AFF throughput estimation algorithm}
 \begin{algorithmic}[1]
 \renewcommand{\algorithmicrequire}{\textbf{Input:}}
 \renewcommand{\algorithmicensure}{\textbf{Output:}}
 \REQUIRE $maxbitrateindex$, $bitratebw(i)$, \\  $SegmentIndex = n$, $THR_n$ \\ 
 \ENSURE  $switchbitrate(i)$
 \\ \textit{Initialization} : $m_0$ = $w_0$ = 0; $\lambda_0$ = 0; \\ $\Omega_1$ = $\Delta_1$ = 0; $\eta$ =0.1;
 \IF {($SegmentIndex == 1$)}
  \STATE $m_1=(\lambda_0* m_0)+THR_1$
  \STATE $w_1=(\lambda_0* w_0)+1$
  \STATE $\overline{THR_1}=\frac{m_1}{w_1}$
  \STATE $\lambda_{1}=\lambda_0-\eta*2*(\overline{THR_1}-THR_1)*(\frac{\Delta_1*w_1-\Omega_1*m_1}{w_1^2})$
  \ELSE
  \STATE $\Delta_{n}=(\lambda_{n-1}*\Delta_{n-1})+m$
  \STATE $\Omega_{n}=(\lambda_{n-1}*\Omega_{n-1})+w$
  \STATE $m_{n}=(\lambda_{n-1}*m_{n-1})+THR_n$
  \STATE $w_{n}=(\lambda_{n-1}*w_{n-1})+1$
  \STATE $\overline{THR_n}=\frac{m_{n}}{w_{n}}$
  \STATE $\lambda_{n}=\lambda_{n-1}-\eta*2*(\overline{THR_{n}}-THR_{n})*(\frac{\Delta_{n}*w_{n}-\Omega_{n}*m_{n}}{w_{n}^2})$
  \ENDIF
  \FOR {$i = maxbitrateindex$ to $0$}
  \IF {(${\overline{THR_n}} \geq bitratebw(i)$)}
  \RETURN $switchbitrate(i)$
  \STATE \textbf{end for} 
  \ENDIF
  \ENDFOR
 \end{algorithmic} 
 \end{algorithm}

The AFF algorithm is executed every time a new instant throughput measurement is obtained, i.e., every time a video segment is downloaded. Once the average throughput is estimated, the algorithm selects a new video representation by performing a sequential search on the video bitrates of the different representations, from highest to lowest, to select the highest bitrate that is below the average throughput. \par
Bodenham in \cite{bodenhamt}, performing numerous simulations, reached the conclusion that truncating the range of $\overrightarrow{\lambda}$ to $\overrightarrow{\lambda}\in(0.6,1)$ can make the method react faster when encountering short-term fluctuations.\par
Bodenham also defined that the step size $\eta$ variable should be between $\eta\in(0.001,0.1)$ so that the algorithm can return to a steady behavior faster when detecting a fluctuation. In this proposal a value of $\eta=0.1$ is used. \par

The next section presents the scenario in which the AFF adaptive algorithm was evaluated.\par

\section{Implementation Scenario}
To test the proposal, a DASH player has been modified to add AFF as the throughput estimation algorithm. The dash.js player \cite{dash} has been selected. This player is a JavaScript implementation of a DASH player that can run in a web browser. The original dash.js reference player implements a throughput algorithm that consists of calculating the average of the throughput obtained for the last three video segments. This algorithm will be referred to as \textit{avg-last-3}. The dash.js player also implements a buffer control algorithm so that it can adapt differently when certain established levels are reached. For instance, when the buffer level drops to 8 seconds, lower bitrate segments are requested independently of how much bandwidth the throughput algorithm estimates. \par 

Figure 1 presents a block diagram of the adaptation logic implemented in dash.js. There are four elements in the adaptation algorithm. From those four elements, the complexity of the adaptation logic resides mainly on the Buffer Module and the Throughput Module, which are the modules implementing the algorithms managing the selection of the next segment bitrate.\par 

The Rules Entity defines all the necessary parameters and levels such as the minimum buffer level or if the streaming is live or pre-stored. The Throughput Module is where the instant throughput is measured and where the avg-last-3 method to estimate throughput is implemented. This module has been modified to implement the AFF method and the EWMA method so the three different throughput estimation methods can be compared. \par
The Buffer Module implements a heuristic algorithm that consists of monitoring the buffer level to force a bitrate change when the buffer is under a minimum level. Finally, the Switching Logic controls the request for new video segments choosing a video representation bitrate that fits the conditions coming from the Buffer Module and the Throughput Module.\par

Our experimental platform consists of a client-server model which is interconnected by a network emulator, as shown in Fig. 3. The scenario has been implemented using an open-source virtualized platform \cite{lentisco}.\par

\begin{figure}[!t]
\centering
    \includegraphics[width=3.1in]{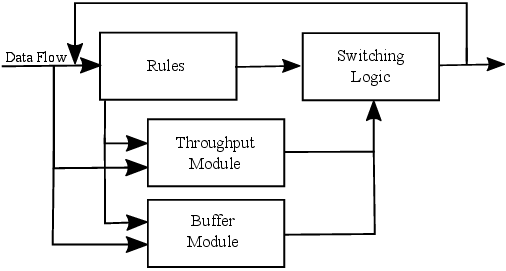}
\caption{Block diagram of the adaptation logic of dash.js.}
\label{fig1:verticalcell}
\end{figure}\par
\begin{figure}[!t]
\centering
    \includegraphics[width=3.05in]{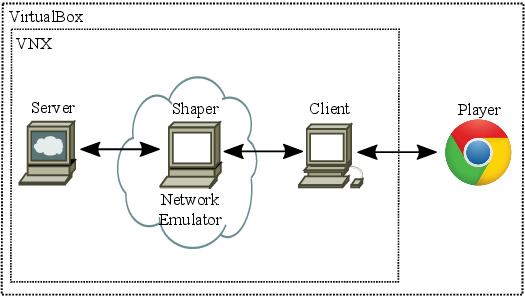}
\caption{Experiment scenario.}
\label{fig2:verticalcell}
\end{figure}\par

This scenario is built over two layers of virtualization; the first layer consists of an open-source tool named Virtual Networks over linuX (VNX) \cite{vnx}. This tool is based on Linux Containers (LXC) and an XML file to create virtual network scenarios. The XML file contains the information on the number of network elements, the interconnections among them and commands to perform specific tasks. The second layer of virtualization consists of hosting the VNX scenario in a virtual machine running Ubuntu inside VirtualBox \cite{vb}. This layer provides the portability for the scenario to run in any OS with the use of an Open Virtual Appliance (OVA).\par
The server stores the Big Buck Bunny video \cite{bbb} at a resolution of 720p. The video is encoded in H.264/Advanced Video Coding (AVC) format with variable bitrate (VBR) using the x264 tool \cite{x264} with four different bitrates (250, 500, 1000 and 2000 Kbps). MP4Box \cite{gpac} is used to split each video file in segments and generate the MPD file. Each video segment has a duration of 2 seconds.\par
The shaper behaves as a network emulator that modifies the network conditions in three aspects: available bandwidth, network delay and packet loss. The shaper is based on a shell script that reads a Comma-separated Value (CSV) network profile file defining the network conditions on different time periods. This file is used to change the configuration of network interfaces during the video streaming experiments, using the traffic control and the network emulation (NetEm) tools of Linux. More specifically, the shaper uses the Hierarchy Token Bucket (HTB) queuing discipline to classify the incoming TCP traffic and enforce the bandwidth parameter read from the network profile file. \par
The client has a caching proxy, that can be used to support broadcast streaming services. In our scenario, as depicted in Fig. 2, the player consists of a Chrome web client running dash.js to access the video segments transparently using the client proxy. The player is also responsible for collecting the data which is used to analyze the behavior of the throughput estimation algorithm; i.e., buffer length, video bitrate and instant throughput.\par
The AFF throughput estimation method has been compared to the dash.js avg-last-3 estimation method \cite{dash} and to our own implementation of the EWMA method described in \cite{akh,thang2}. For EWMA, the fixed weight parameter is set to 0.2, which is the value for wireless network proposed by \cite{akh}, since the profiles are based on LTE traffic measurements.\par
The experiments are carried out using four different bandwidth profiles, as shown in Fig. 4. The three first profiles use real LTE network data obtained in a previous research \cite{lentisco}. The fourth profile was created to analyze the behavior of the throughput estimation methods when the available bandwidth changes abruptly. \par

\begin{figure}[!t]
\centering
    \includegraphics[width=3.1in]{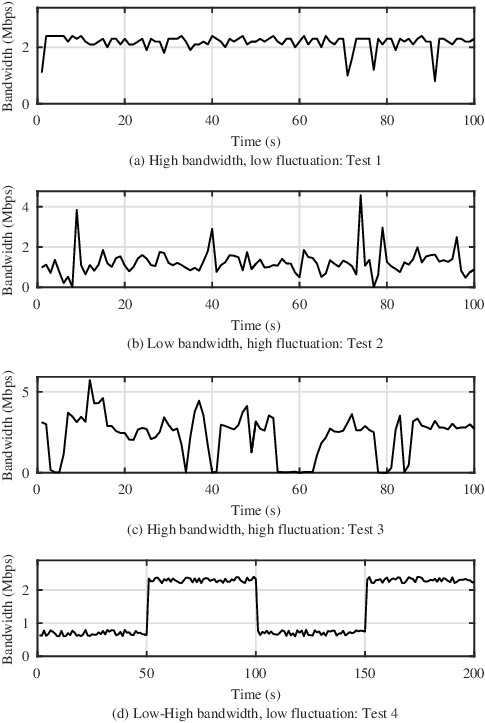}
\caption{Bandwidth profiles used.}
\label{fig3:verticalcell}
\end{figure}

\begin{table}[!t]
\renewcommand{\arraystretch}{1.3}
\caption{Statistics of the bandwidth profiles in Mbps}
\label{tab:profiles1}
\centering
\begin{tabular}{cccccc}
\toprule
 \textbf{Test} & \textbf{Bandwidth Type} & \textbf{Max.} & \textbf{Min.} & \textbf{Avg.} & \textbf{S.D.} \\
\midrule
Test 1 & \multicolumn{1}{p{2.2cm}}{\centering High bandwidth, low fluctuation.} & 2.40 & 0.80 & 2.17 & 0.2765\\
Test 2 & \multicolumn{1}{p{2.2cm}}{\centering Low bandwidth, High fluctuation.} & 4.57 & 0.01 & 1.23 & 0.6374\\
Test 3 & \multicolumn{1}{p{2.2cm}}{\centering High bandwidth, High fluctuation.} & 5.73 & 0.01 & 2.31 & 1.3317\\
Test 4 & \multicolumn{1}{p{2.2cm}}{\centering High-low bandwidth, low fluctuation.} & 2.39 & 0.60 & 1.50 & 0.8063\\
\bottomrule
\end{tabular}
\end{table}

The first bandwidth profile was selected to represent the behavior of a steady network such as a residential Internet connection. The second profile aims to study the performance of the throughput estimation methods for limited bandwidths that present short-term fluctuations. The third profile is inspired by test pattern one (TP1) in \cite{kim}, depicting a user walking during daytime in an urban environment. The fourth profile was created to analyze the behavior of the different throughput estimation methods when a sudden loss of bandwidth occurs, especially to measure how a high change in the bandwidth affect the algorithms that carry past measurements when estimating the available throughput.\par

Table \ref{tab:profiles1} shows the main statistics for each of the bandwidth profiles used to test the performance of the AFF algorithm: maximum, minimum, average, and standard deviation for the bandwidth of each profile.\par

\section{Performance Evaluation}

In this section we present the experimental results of evaluating the different throughput estimation methods avg-last-3, EWMA and AFF for each of the bandwidth profiles. \par
The first subsection presents an analysis of the three different throughput estimation methods using the first bandwidth profile, and the behavior of $\lambda$ for the AFF method. The next four subsections present the results for each bandwidth profile in the form of a Cumulative Distribution Function (CDF) for the video bitrate selection and the buffer level obtained for the three throughput estimation methods. Finally, the last subsection discusses the Quality of Experience (QoE) indicators for each of the scenarios that are summarized in Table \ref{tab:statistics}.\par

\subsection{Comparison of throughput estimation methods}
\begin{figure}[!t]
\centering
    \includegraphics[width=3.1in]{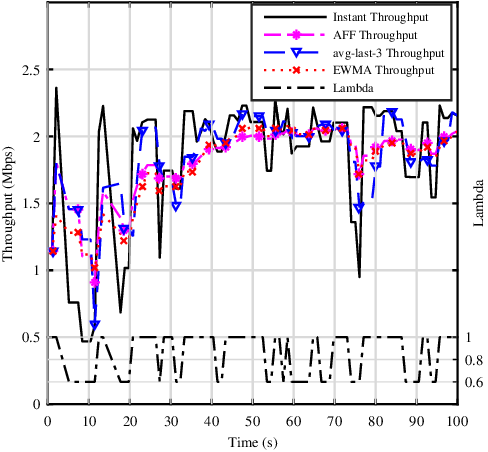}
\caption{Estimated throughput of the methods and lambda in bandwidth profile Test 1.}
\label{fig4:verticalcell}
\end{figure}

Fig. 4 shows measurements of the instant throughput that the DASH player calculates, the estimates that each method obtains from the instant throughput measurements, and the lambda value that the AFF method calculates adaptively. Since AFF and EWMA share a similar weighting mechanism, they present a similar behavior, and different from avg-last-3 that just calculates the mean value of the three past measurements. \par
Fig. 4 also shows how the lambda AFF factor evolves during the playback of the video. When the instant throughput behaves in a steady manner, $\lambda=1$, the highest value. But when there is a short-term fluctuation of the throughput, lambda drops to the lowest value of $\lambda =0.6$, which makes the AFF algorithm to forget the latest measurements faster.\par
\subsection{High available bandwidth with low short-term fluctuations}
This test was conducted to compare the three algorithms in a steady environment with enough bandwidth to reach the highest representation most of the time. This type of bandwidth profile describes the typical bandwidth of a residential Internet connection and it is similar to test pattern two (TP2) in \cite{kim}.
\par

\begin{figure}[!t]
\centering
    \includegraphics[width=3.1in]{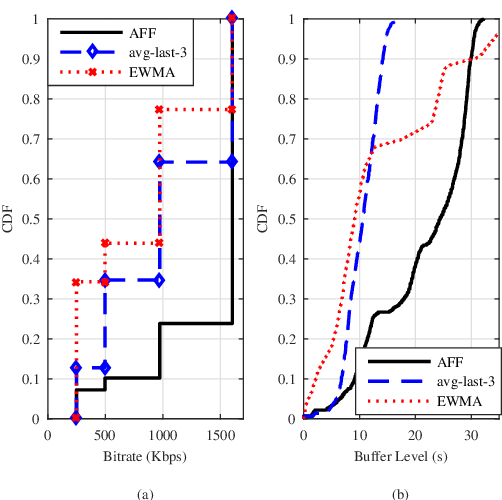}
\caption{(a) CDF of bitrate and (b) CDF of buffer level in Test 1.}
\label{fig5:verticalcell}
\end{figure}
Fig. 5 presents the CDF of the bitrate selection for each of the algorithms, showing that the AFF algorithm is more stable than the other algorithms since the best quality is chosen most of the time.\par
Fig. 5 also presents the CDF of the buffer level. The AFF algorithm maintains a buffer level above ten seconds, which allows the AFF throughput estimation method to choose the bitrates instead of the buffer control algorithm. It also shows that the avg-last-3 and EWMA methods present more bitrate changes.
\subsection{Low available bandwidth with high short-term fluctuations}
In the second test, a low bandwidth and high number of short-term bandwidth fluctuations profile is used. This profile affects video quality because of the sudden drops of available bandwidth. It may also cause stalling during the playback because of the continuous fluctuation in bandwidth. This profile was selected to evaluate the behavior of the different throughput estimation methods in an environment where the bandwidth is low.
\par
\begin{figure}[!t]
\centering
    \includegraphics[width=3.1in]{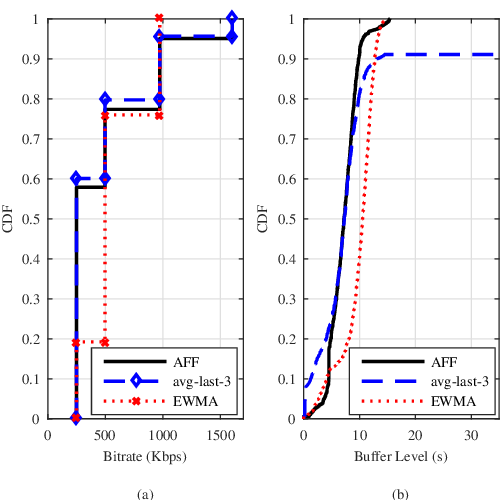}
\caption{(a) CDF of bitrate and (b) CDF of buffer level in Test 2.}
\label{fig6:verticalcell}
\end{figure}
Fig. 6 shows that in this case the EWMA method presents better results than the AFF method and avg-last-3 method. Both the AFF and the avg-last-3 methods select the lowest bitrate quality most of the playback time. However, the dahs.js method present a stalling event of five seconds that affects the quality of the playback. The buffer level behaved in a similar manner with the three methods, mostly under 10 seconds because of the low bandwidth. 
\par

\subsection{High available bandwidth with high short-term fluctuations}
The high bandwidth and high short-term fluctuations of this profile are expected to show whether the buffer can minimize the impact of having severe drops from high bandwidth measurements, as well as the behavior of the different throughput estimation methods. This profile represents the bandwidth behavior of an LTE user who is walking in an urban environment, and it is similar to TP1 in \cite{kim}.
\par

As shown in Fig. 7, since the bandwidth profile has many high short-term fluctuations with high bandwidth, the AFF and the avg-last-3 methods selects the highest bitrate most of the time. However, the AFF method selects the highest bitrate quality ten percent more than the avg-last-3 method. The EWMA method shows a poor behavior, selecting the third bitrate quality most of the time. The buffer level behaves differently for each method; however, it is shown that the EWMA method presented the most bitrate switches because the buffer level stays below ten seconds most of the time, which triggers the buffer level mechanism. The results from the first and third tests show that the AFF method is suited to perform proficiently in wired and wireless environments.
\par
\begin{figure}[!t]
\centering
    \includegraphics[width=3.1in]{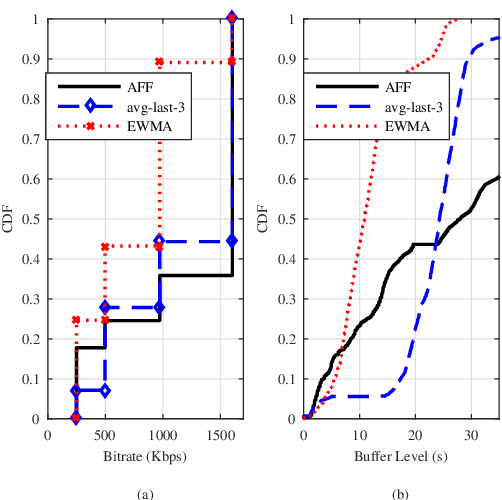}
\caption{(a) CDF of bitrate and (b) CDF of buffer level in Test 3.}
\label{fig7:verticalcell}
\end{figure}
\begin{figure}[!t]
\centering
    \includegraphics[width=3.1in]{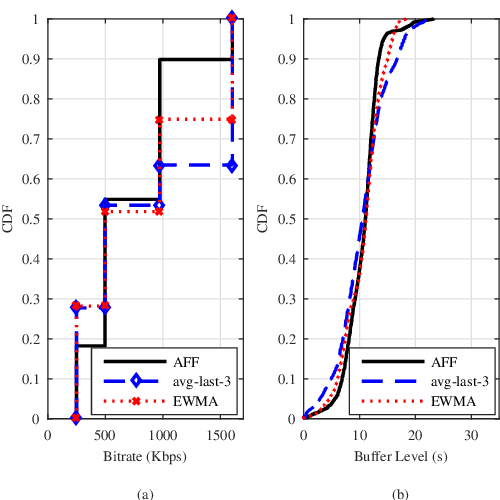}
\caption{(a) CDF of bitrate and (b) CDF of buffer level in Test 4. }
\label{fig8:verticalcell}
\end{figure}
\begin{table*}
\centering
\caption{Statistics of the Algorithms}
\label{tab:statistics}
\begin{tabular}{ccccccc}
\toprule
\textbf{Test} & \textbf{Method} & \textbf{Bitrate Changes} & \textbf{Stalling Events} & \textbf{Time (s)} & \textbf{Mean Bitrate (Kbps)}\\
\midrule
  \multirow{3}{*}{Test 1} & AFF & \textbf{7} & 0 & -- & \textbf{1387.00} \\
 & avg-last-3 & 41 & 0 & -- & 1002.10 \\
 & EWMA & 21 & \textbf{2} & 1.28, 0.9 & 822.12 \\
 \\
   \multirow{3}{*}{Test 2} & AFF & 34 & 0 & -- & 492.00 \\
 & avg-last-3 & 35 & \textbf{1} & 5.01 & 471.83 \\
 & EWMA & \textbf{14} & 0 & -- & \textbf{563.00} \\
 \\
   \multirow{3}{*}{Test 3} & AFF & \textbf{8}& 0 & -- & \textbf{1216.80} \\
 & avg-last-3 & 15 & 0 & -- & 1174.00 \\
 & EWMA & 25 & 0 & -- & 775.00 \\
 \\
   \multirow{3}{*}{Test 4} & AFF & \textbf{26} & 0 & -- & 730.00 \\
 & avg-last-3 & 33 & \textbf{2} & 0.3, 2.23 & \textbf{880.00} \\
 & EWMA & 30 & 0 & -- & 814.00 \\
\bottomrule
\end{tabular}
\end{table*}
\subsection{Low and high bandwidth variations}

This profile combines low bandwidth and high bandwidth with low number of short-term fluctuations. This profile was created to analyze the impact of older bandwidth estimations on the accuracy of the throughput estimation methods when there is an abrupt bandwidth change. \par
Fig. 8 shows that the AFF method selects the second and third bitrate quality most of the time, and the smallest number of quality changes. The avg-last-3 method reaches the highest mean bitrate, but suffers from two stalling events of 0.3 seconds and 2 seconds respectively, as Table \ref{tab:statistics} shows. The EWMA method presents a better mean bitrate than the AFF method but selects the lowest quality more often and presents more bitrate quality switches.
 It also shows that the buffer level is lower than ten seconds half of the time, which means that the buffer control algorithm affects the throughput estimation by constantly lowering the bitrate chosen for this bandwidth profile.\par
\par
\subsection{QoE Results}

Table \ref{tab:statistics} shows, for each of the methods, a set of quality of experience indicators: number of bitrate changes, number of stalling events (due to buffer under-runs), the duration of each stalling event in seconds and the mean bitrate in Kbps. In the first test the AFF method presented a better-quality performance since there were less bitrate changes and the mean bitrate was higher than for the other methods. It also shows that the EWMA method presented two stalling events that lasted 1.28 and 0.9 seconds respectively, having a great impact on the QoE of the playback. In the second bandwidth profile, the EWMA method presented a steadier behavior than the other two methods; however, the mean bitrate of each method is not far from one another. It also shows that the avg-last-3 method presented a stalling event that lasted 5 seconds. For the third bandwidth profile, the results show that the AFF method presents less bitrate changes and a better mean bitrate.
 In the fourth test, the AFF method presents less bitrate changes, however, the avg-last-3 method had the best mean bitrate. The problem with the avg-last-3 method is that it had two stalling events, one lasted three seconds and the other one two seconds, making this method not suited to be used in environments with significant drops in bandwidth. \par

\section{Fairness analysis}
\begin{figure}[!t]
\centering
    \includegraphics[width=3.1in]{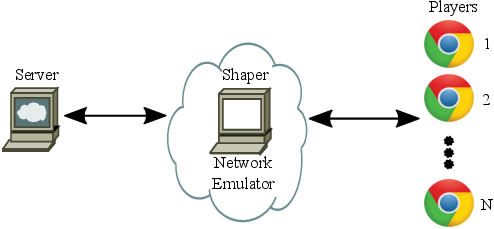}
\caption{Fairness experiment scenario.}
\label{fig9:verticalcell}
\end{figure}
In adaptive video streaming, fairness metrics are used to determine if the adaptation algorithm is able to deliver a fair share of the network bandwidth to different clients. In order to analyze the fairness of the AFF algorithm we have created an experimental scenario, as shown in Fig. 9\label{fig9}, to measure the bandwidth obtained by several simultaneous players sharing a bottleneck link. \par

This scenario is similar to the scenario explained in Section IV, but in this case the equipment is not virtualized. The HTTP server, the multimedia content and the multimedia player are described in Section IV. The shaper is now implemented using a Raspberry Pi 3 with two network interfaces. The shaper uses a network profile as specified in Table \ref{tab:profiles2}. This profile was created to measure the fairness of the AFF estimation algorithm and the avg-last-3 algorithm, in order to detect how the algorithm responds to changes in the bandwidth when competing with other video players using the same algorithm. \par

\begin{table}[!t]
	\renewcommand{\arraystretch}{1.3}
	\caption{Fairness analysis network profile}
	\label{tab:profiles2}
	\centering
	\begin{tabular}{cc}
		\toprule
		\textbf{Interval (s)} & \textbf{Bandwidth (Mbps) }\\
		\midrule
		0-100  & 22\\
		100-200  & 12\\
		200-300  & 6\\
		300-360  & 22\\				
		\bottomrule
	\end{tabular}
\end{table}

The experiments consist of ten players initiating the multimedia playback randomly within the first 15 s of the experiment. The results are obtained measuring the average bandwidth of each client in the interval 50-350 s. These values are used to calculate the Jain Fair index (JFI) \cite{jain} and the total average throughput, as shown in Table \ref{tab:fairness}. \par

\begin{table}[!t]
\renewcommand{\arraystretch}{1.3}
\caption{Fairness analysis results}
\label{tab:fairness}
\centering
\begin{tabular}{cccc}
\toprule
\textbf{Test} & \textbf{Algorithm } & \textbf{JFI} & \textbf{Avg. Thr. (Kbps) }\\
\midrule
Test 5  & AFF & .9969 & \textbf{1281.7}\\
Test 6  & avg-last-3 & \textbf{.9995} & 1188.5\\

\bottomrule
\end{tabular}
\end{table}
The results show that both algorithms, with a JFI value close to 1, present a fair use of the bandwidth. Thus, the AFF estimation algorithm shows fairness results that are similar to already existing algorithms, making the AFF algorithm suited to be used as the throughput estimation mechanism in combination with different adaptation algorithms. The results also show that the AFF method manages to achieve a slightly higher average throughput for the clients.\par

\section{Conclusion and Future Work}

Adaptive bitrate streaming solutions rely on throughput estimation algorithms that might need fine-tuning depending on the characteristics of the network. In this paper, we propose to use the Adaptive Forgetting Factor method for throughput estimation in DASH adaptation algorithms. This method relies on instant throughput measurements that are aggregated using exponential weights adaptively, so it overcomes the problems of short-term fluctuations in network throughput. By using this method it is possible to improve the QoE obtained by a DASH player over different types of networks, avoiding the fine-tuning of parameters of other throughput estimation algorithms. Using different bandwidth profiles, the AFF proposal has been tested and compared to alternative throughput estimation methods such as EWMA and average-last-3. The results show that AFF might obtain slightly lower average bitrates than the alternative methods. However, it presents a better behavior in regard to video stalling and number of bitrate switches, which are two of the key parameters for QoE in adaptive streaming. Finally, the results of the fairness experiments show that the AFF algorithm is able to deliver a fair share of the network bandwidth to different clients. Therefore, we propose AFF as a valid throughput estimation method that can work adequately for DASH players over different types of networks.\par
A future step in our research is to work on DASH adaptation algorithms that can exploit a better communication between throughput estimation and buffer control methods, to provide the best QoE with different network conditions. \par


%




\ifCLASSOPTIONcaptionsoff
  \newpage
\fi



%
\bibliographystyle{IEEEtran}
\bibliography{references}

%



\begin{IEEEbiography}
[{\includegraphics[width=1in,height=1.25in,clip,keepaspectratio]{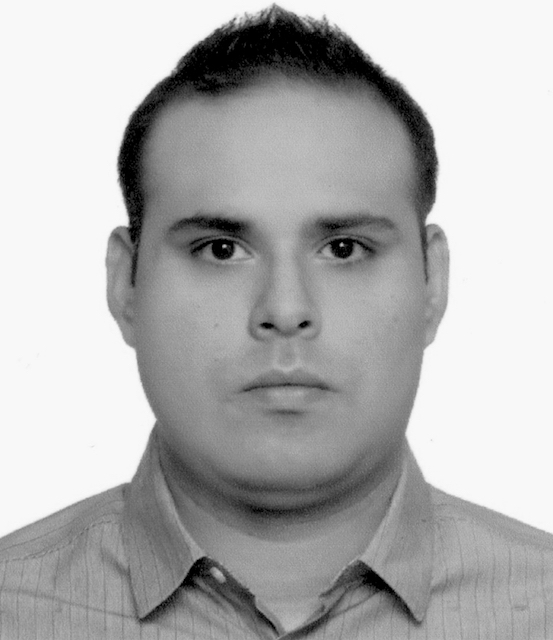}}]
{Miguel Aguayo}
received the B.E. degree in Telematic Engineering from the Universidad de Colima, Colima, M\'exico, in 2002 and the M.S. degree in electronics and telecommunications from the Ensenada Center for Scientific Research and Higher Education (CICESE), Ensenada, M\'exico, in 2004. Currently, he is a Ph.D. candidate in the Department of Telematics Systems Engineering at the Universidad Polit\'ecnica de Madrid (UPM). His research interests include multimedia communications, internetworking, mobile networks, and quality of experience.
\end{IEEEbiography}

\begin{IEEEbiography}
[{\includegraphics[width=1in,height=1.25in,clip,keepaspectratio]{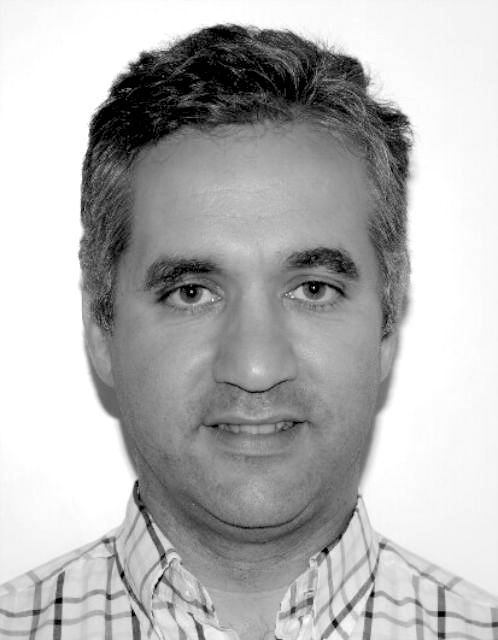}}]{Luis Bellido}
received the M.S. and Ph.D. degrees in Telecommunications Engineering from the Universidad Polit\'ecnica de Madrid (UPM), Spain, in 1994 and 2004, respectively. He is currently an Associate Professor at UPM, specializing in the fields of computer networking, {Internet} technologies and quality of service. His current research interests include mobile networks, multimedia applications, and virtualization.
\end{IEEEbiography}

\begin{IEEEbiography}
[{\includegraphics[width=1in,height=1.25in,clip,keepaspectratio]{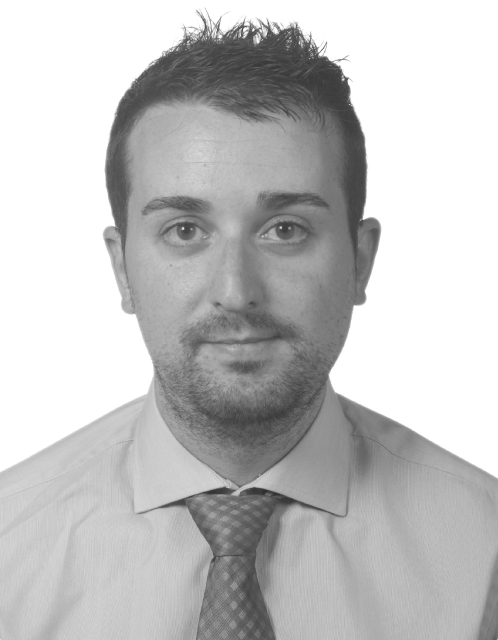}}]
{Carlos M. Lentisco}
received the Telecommunications Engineering degree (2013) from the Universidad Carlos III de Madrid and the M.S. degree (2014) in Networks and Telematic Services Engineering from Universidad Polit\'ecnica de Madrid (UPM). Currently, he is a Ph.D. candidate in Telematics Systems Engineering Department, UPM. His research interests include multimedia streaming, mobile networks, and virtualization.
\end{IEEEbiography}

\begin{IEEEbiography}
[{\includegraphics[width=1in,height=1.25in,clip,keepaspectratio]{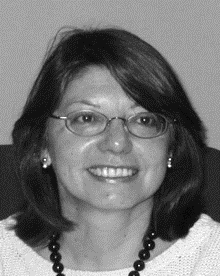}}]{Encarna Pastor}
received the M.S. and Ph.D. degrees in Computer Science from the Universidad Polit\'ecnica de Madrid (UPM) in 1980 and 1988, respectively. She is currently a Full Professor at UPM, specializing in the fields of computer networking, multimedia applications, and Internet technologies. Her current research interests include content delivery networks, multimedia networking, and quality of experience.
\end{IEEEbiography}

\vfill





\end{document}